\documentclass[a4paper,aps,twocolumn,nofootinbib]{revtex4}
\RequirePackage[colorlinks,hyperindex]{hyperref}
\RequirePackage[english]{babel}
\RequirePackage[latin1]{inputenc}
\RequirePackage[T1]{fontenc}
\RequirePackage{mathrsfs}
\RequirePackage{amsmath}
\RequirePackage{amssymb}
\RequirePackage{amsbsy}
\RequirePackage{color}
\RequirePackage{bm}
\hypersetup{colorlinks=true,breaklinks=true,urlcolor=blue,linkcolor=red}
\begin{document}
\title{\bf{The most complete mass-dimension four topological gravity}}
\author{Luca Fabbri}
\affiliation{DIME, Sez. Metodi e Modelli Matematici, Universit\`{a} di Genova,\\
Via all'Opera Pia 15, 16145 Genova, ITALY}
\date{\today}
\begin{abstract}
The usual Chern-Simons extension of Einstein gravity theory consists in adding a squared Riemann contribution to the Hilbert Lagrangian, which means that a square-curvature term is added to the linear-curvature leading term governing the dynamics of the gravitational field. However, in such a way the Lagrangian consists of two terms with a different number of curvatures, and therefore not homogeneous. To develop a homogeneous Chern-Simons correction to Einstein gravity we may, on the one hand, use the above-mentioned square-curvature contribution as the correction for the most general square-curvature Lagrangian, or on the other hand, find some linear-curvature correction to the Hilbert Lagrangian. In the first case, we will present the most general square-curvature leading term, which is in fact the already-known re-normalizable Stelle Lagrangian. In the second case, the topological current has to be an axial-vector built only in terms of gravitational degrees of freedom and with a unitary mass dimension, and we will display such an object. The comparison of the two theories will eventually be commented.
\end{abstract}
\maketitle
\section{Introduction}
Einstein gravity is perhaps one of the best theories ever to be built. Constructed starting from first principles of the most general validity, it has passed all the experimental tests performed in astrophysics and cosmology.

As a matter of fact, if dark matter is indeed, as it seems to be, some form of matter, and not a modification of the gravitational field, then there is not a single observation that Einsteinian gravitation would not fit. From a purely theoretical perspective, Einstein gravity has always been thought to be doomed by the necessity of singularity formation, but this is not actually a problem either, as the formation of singularity is no longer an unavoidable feature of Einstein gravity if it is complemented by torsion and taken in presence of Dirac spinor matter fields \cite{Fabbri:2017rjf,Fabbri:2017xch}.

Just the same, one might wonder, again under a purely theoretical point of view, what would happen if topological contributions were to be included in this gravitational theory. The earliest attempt to include a Chern-Simons type of term into Einsteinian gravitation is that of \cite{Jackiw:2003pm} (for a review see \cite{Alexander:2009tp} and references therein), and it consists in adding one specific squared Riemann contribution to the Hilbert Lagrangian. The motivation for adding a term of type $R^{pq\sigma\nu}R^{ac}_{\phantom{ac}\sigma\nu}\varepsilon_{pqac}$ is in analogy with the $F^{pq}F^{ac}\varepsilon_{pqac}$ found in electrodynamics, but while this can mathematically be done, nevertheless one cannot help but notice a certain lack of homogeneity. In fact, in electrodynamics the leading term has the structure $F^{ac}F_{ac}$ and therefore gravity should have a leading term $R^{pq\sigma\nu}R_{pq\sigma\nu}$ to obtain a full analogy. Such a square-curvature term, or a closely related one, is already studied by Stelle in \cite{Stelle:1977ry, Stelle:1976gc}. However the complementary way is to add $R^{pqac}\varepsilon_{pqac}$ to the term $R$ that constitutes the standard Hilbert Lagrangian. The problem with this term is that it is zero, and therefore a term that is a pseudo-scalar with $2$ mass dimension must be found in alternative. This amounts to ask that such a topological correction be of the form $\nabla_{\nu}K^{\nu}$ with $K^{\nu}$ an axial-vector of unitary mass dimension. Or equivalently, $K^{\nu}$ must be an axial-vector built in terms of the connection. Although there seems to be no such a thing, recent studies are useful in providing this topological current as we are going to discuss in the present paper.
\section{Spinor Fields}
As anticipated in the introduction, recent findings have enabled us to obtain a topological current $K^{\nu}$ constructed in terms of the connection alone. This can be done when Dirac spinorial matter fields are the space-time content.

The Clifford matrices $\boldsymbol{\gamma}^{a}$ are defined as $\left\{\boldsymbol{\gamma}_{a}\!,\!\boldsymbol{\gamma}_{b}\right\}\!=\!2\eta_{ab}\mathbb{I}$ where $\eta_{ab}$ is the Minkowskian matrix. So $\left[\boldsymbol{\gamma}_{a}\!,\!\boldsymbol{\gamma}_{b}\right]\!=\!4\boldsymbol{\sigma}_{ab}$ defines the generators of the complex Lorentz algebra and the relationship $2i\boldsymbol{\sigma}_{ab}\!=\!\varepsilon_{abcd}\boldsymbol{\pi}\boldsymbol{\sigma}^{cd}$ defines $\boldsymbol{\pi}$ (this matrix is usually denoted as a gamma matrix with an index five, but in space-time this index has no meaning, and so we use a notation with no index). By exponentiating $\boldsymbol{\sigma}_{ab}$ we can compute the local complex Lorentz group $\boldsymbol{S}$ and the spinor field $\psi$ is defined as what transforms according to $\psi\!\rightarrow\!\boldsymbol{S}\psi$ in general. With the Clifford matrices we define the adjoint spinor $\overline{\psi}\!=\!\psi^{\dagger}\boldsymbol{\gamma}^{0}$ again in general. The set
\begin{eqnarray}
&\Sigma^{ab}\!=\!2\overline{\psi}\boldsymbol{\sigma}^{ab}\boldsymbol{\pi}\psi\\
&M^{ab}\!=\!2i\overline{\psi}\boldsymbol{\sigma}^{ab}\psi
\end{eqnarray}
\begin{eqnarray}
&S^{a}\!=\!\overline{\psi}\boldsymbol{\gamma}^{a}\boldsymbol{\pi}\psi\\
&U^{a}\!=\!\overline{\psi}\boldsymbol{\gamma}^{a}\psi
\end{eqnarray}
\begin{eqnarray}
&\Theta\!=\!i\overline{\psi}\boldsymbol{\pi}\psi\\
&\Phi\!=\!\overline{\psi}\psi
\end{eqnarray}
defines bi-linear spinor quantities, they are all real and such that $U_{a}U^{a}\!=\!-S_{a}S^{a}\!=\!|\Theta|^{2}\!+\!|\Phi|^{2}$ and $U_{a}S^{a}\!=\!0$ hold.

These bi-linears can be used to perform the Lounesto classification \cite{L, Cavalcanti:2014wia}: singular spinors are those for which $\Theta\!=\!\Phi\!\equiv\!0$ \cite{HoffdaSilva:2017waf, daSilva:2012wp, Ablamowicz:2014rpa, daRocha:2013qhu}; regular spinors are defined when either $\Theta$ or $\Phi$ or both are not equal to zero identically. In this paper, we will be interested in regular spinors, and that is Dirac spinors, for which it is always possible to write
\begin{eqnarray}
&\!\psi\!=\!\phi e^{-\frac{i}{2}\beta\boldsymbol{\pi}}
\boldsymbol{S}\left(\!\begin{tabular}{c}
$1$\\
$0$\\
$1$\\
$0$
\end{tabular}\!\right)
\label{spinor}
\end{eqnarray}
in chiral representation, for some complex Lorentz transformation $\boldsymbol{S}$, with $\beta$ and $\phi$ called Yvon-Takabayashi angle and module, and where the spinor is said to be in polar form \cite{Fabbri:2016msm}. The bi-linear spinor quantities reduce to
\begin{eqnarray}
&\Sigma^{ab}\!=\!2\phi^{2}(\cos{\beta}u^{[a}s^{b]}\!-\!\sin{\beta}u_{j}s_{k}\varepsilon^{jkab})\\
&M^{ab}\!=\!2\phi^{2}(\cos{\beta}u_{j}s_{k}\varepsilon^{jkab}\!+\!\sin{\beta}u^{[a}s^{b]})
\end{eqnarray}
showing that they can always be written with the vectors
\begin{eqnarray}
&S^{a}\!=\!2\phi^{2}s^{a}\\
&U^{a}\!=\!2\phi^{2}u^{a}
\end{eqnarray}
and the scalars
\begin{eqnarray}
&\Theta\!=\!2\phi^{2}\sin{\beta}\\
&\Phi\!=\!2\phi^{2}\cos{\beta}
\end{eqnarray}
such that $u_{a}u^{a}\!=\!-s_{a}s^{a}\!=\!1$ and $u_{a}s^{a}\!=\!0$ and which show that Yvon-Takabayashi angle and module are the only $2$ true degrees of freedom. The $8$ real components of spinors are rearranged into the special configuration in which the $2$ real scalar degrees of freedom, YT angle and module, become isolated from the $6$ components that can always be transferred away, the spin and the velocity. We notice that the YT angle is a zero-dimension pseudo-scalar and therefore the module inherits the full $3/2$-dimension that characterizes the spinor field. This is one most important remark for the following of the paper, as we shall see.

As for the background, we have that using the metric we define the symmetric connection $\Lambda^{\sigma}_{\alpha\nu}$ and with it we define the spin connection $\Omega^{a}_{\phantom{a}b\pi}\!=\!\xi^{\nu}_{b}\xi^{a}_{\sigma}(\Lambda^{\sigma}_{\nu\pi}\!-\!\xi^{\sigma}_{i}\partial_{\pi}\xi_{\nu}^{i})$ and
\begin{eqnarray}
&\boldsymbol{\Omega}_{\mu}
=\frac{1}{2}\Omega^{ab}_{\phantom{ab}\mu}\boldsymbol{\sigma}_{ab}
\!+\!iqA_{\mu}\boldsymbol{\mathbb{I}}\label{spinorialconnection}
\end{eqnarray}
in terms of the gauge potential $qA_{\mu}$ and called spinorial connection. This is needed to write
\begin{eqnarray}
&\boldsymbol{\nabla}_{\mu}\psi\!=\!\partial_{\mu}\psi
\!+\!\boldsymbol{\Omega}_{\mu}\psi\label{spincovder}
\end{eqnarray}
as spinorial covariant derivative. Then the commutator of spinorial covariant derivatives justifies the definitions
\begin{eqnarray}
&R^{i}_{\phantom{i}j\mu\nu}\!=\!\partial_{\mu}\Omega^{i}_{\phantom{i}j\nu}
\!-\!\partial_{\nu}\Omega^{i}_{\phantom{i}j\mu}
\!+\!\Omega^{i}_{\phantom{i}k\mu}\Omega^{k}_{\phantom{k}j\nu}
\!-\!\Omega^{i}_{\phantom{i}k\nu}\Omega^{k}_{\phantom{k}j\mu}\\
&F_{\mu\nu}\!=\!\partial_{\mu}A_{\nu}\!-\!\partial_{\nu}A_{\mu}
\end{eqnarray}
that is the Riemann curvature and the Maxwell strength.

When the polar form is taken into account, and considering that we can formally write the expansion
\begin{eqnarray}
&\boldsymbol{S}\partial_{\mu}\boldsymbol{S}^{-1}\!=\!i\partial_{\mu}\alpha\mathbb{I}
\!+\!\frac{1}{2}\partial_{\mu}\theta_{ij}\boldsymbol{\sigma}^{ij}\label{spintrans}
\end{eqnarray}
we can define
\begin{eqnarray}
&\partial_{\mu}\alpha\!-\!qA_{\mu}\!\equiv\!P_{\mu}\label{P}\\
&\partial_{\mu}\theta_{ij}\!-\!\Omega_{ij\mu}\!\equiv\!R_{ij\mu}\label{R}
\end{eqnarray}
in which we used (\ref{spinorialconnection}) and which can be demonstrated to be tensors and invariant under the gauge transformation simultaneously. With them we can eventually write
\begin{eqnarray}
&\!\!\!\!\!\!\!\!\boldsymbol{\nabla}_{\mu}\psi\!=\!(-\frac{i}{2}\nabla_{\mu}\beta\boldsymbol{\pi}
\!+\!\nabla_{\mu}\ln{\phi}\mathbb{I}
\!-\!iP_{\mu}\mathbb{I}\!-\!\frac{1}{2}R_{ij\mu}\boldsymbol{\sigma}^{ij})\psi
\label{decspinder}
\end{eqnarray}
as spinorial covariant derivative so that
\begin{eqnarray}
&\nabla_{\mu}s_{i}\!=\!R_{ji\mu}s^{j}\label{ds}\\
&\nabla_{\mu}u_{i}\!=\!R_{ji\mu}u^{j}\label{du}
\end{eqnarray}
are general geometric identities. Taking the commutator
\begin{eqnarray}
\!\!\!\!&qF_{\mu\nu}\!=\!-(\nabla_{\mu}P_{\nu}\!-\!\nabla_{\nu}P_{\mu})\label{Maxwell}\\
&\!\!\!\!\!\!\!\!R^{i}_{\phantom{i}j\mu\nu}\!=\!-(\nabla_{\mu}R^{i}_{\phantom{i}j\nu}
\!-\!\!\nabla_{\nu}R^{i}_{\phantom{i}j\mu}
\!\!+\!R^{i}_{\phantom{i}k\mu}R^{k}_{\phantom{k}j\nu}
\!-\!R^{i}_{\phantom{i}k\nu}R^{k}_{\phantom{k}j\mu})\label{Riemann}
\end{eqnarray}
in terms of the Riemann curvature and Maxwell strength, so that they encode electrodynamic and gravitational information. As we have been saying above, in writing the spinor field in its polar form, the spinor field is reconfigured so that its degrees of freedom are isolated from the components transferable into gauge and frames through the $\alpha$ phase and the $\theta_{ij}$ parameters in general. When the phase and parameters are added to gauge potential and spin connection, they do not alter their information, and thus (\ref{P}, \ref{R}) have the information content of the gauge potential and the spin connection themselves, although in the combination non-covariant features fully cancel, so that (\ref{P}, \ref{R}) are gauge invariant and Lorentz covariant, and for this reason they have been called gauge-invariant vector momentum and tensorial connection \cite{Fabbri:2018crr}.

The Dirac spinor dynamics is given by the Lagrangian
\begin{eqnarray}
&\mathscr{L}\!=\!-i\overline{\psi}\boldsymbol{\gamma}^{\mu}\boldsymbol{\nabla}_{\mu}\psi
\!+\!m\overline{\psi}\psi\label{L}
\end{eqnarray}
giving the Dirac spinor field equations
\begin{eqnarray}
&i\boldsymbol{\gamma}^{\mu}\boldsymbol{\nabla}_{\mu}\psi\!-\!m\psi\!=\!0
\label{D}
\end{eqnarray}
as it is well known. Multiplying (\ref{D}) by $\boldsymbol{\gamma}^{a}$ and $\boldsymbol{\gamma}^{a}\boldsymbol{\pi}$ and by $\overline{\psi}$ and splitting real and imaginary parts gives
\begin{eqnarray}
&i(\overline{\psi}\boldsymbol{\nabla}^{\alpha}\psi
\!-\!\boldsymbol{\nabla}^{\alpha}\overline{\psi}\psi)
\!-\!\nabla_{\mu}M^{\mu\alpha}\!-\!2mU^{\alpha}\!=\!0
\label{vr}\\
&\!\!\!\!\nabla_{\alpha}\Phi
\!-\!2(\overline{\psi}\boldsymbol{\sigma}_{\mu\alpha}\!\boldsymbol{\nabla}^{\mu}\psi
\!-\!\boldsymbol{\nabla}^{\mu}\overline{\psi}\boldsymbol{\sigma}_{\mu\alpha}\psi)\!=\!0\label{vi}
\end{eqnarray}
\begin{eqnarray}
&\!\!\!\!\nabla_{\nu}\Theta\!-\!
2i(\overline{\psi}\boldsymbol{\sigma}_{\mu\nu}\boldsymbol{\pi}\boldsymbol{\nabla}^{\mu}\psi\!-\!
\boldsymbol{\nabla}^{\mu}\overline{\psi}\boldsymbol{\sigma}_{\mu\nu}\boldsymbol{\pi}\psi)
\!+\!2mS_{\nu}\!=\!0\label{ar}\\
&(\boldsymbol{\nabla}_{\alpha}\overline{\psi}\boldsymbol{\pi}\psi
\!-\!\overline{\psi}\boldsymbol{\pi}\boldsymbol{\nabla}_{\alpha}\psi)
\!-\!\frac{1}{2}\nabla^{\mu}M^{\rho\sigma}\varepsilon_{\rho\sigma\mu\alpha}\!=\!0\label{ai}
\end{eqnarray}
which are called Gordon decompositions and they have a great importance in writing the Dirac spinor field equation in polar form while maintaining manifest covariance.

Plugging the polar form in (\ref{L}) gives
\begin{eqnarray}
&\mathscr{L}\!=\!-\phi^{2}[s^{\mu}(\nabla_{\mu}\beta\!+\!B_{\mu})
\!+\!2u^{\mu}P_{\mu}\!-\!2m\cos{\beta}]
\end{eqnarray}
while in (\ref{vi}, \ref{ar}) it gives after some manipulation that
\begin{eqnarray}
&\!\!\!\!B_{\mu}\!-\!2P^{\iota}u_{[\iota}s_{\mu]}\!+\!\nabla_{\mu}\beta
\!+\!2s_{\mu}m\cos{\beta}\!=\!0\label{dep1}\\
&\!\!\!\!R_{\mu}\!-\!2P^{\rho}u^{\nu}s^{\alpha}\varepsilon_{\mu\rho\nu\alpha}\!+\!2s_{\mu}m\sin{\beta}
\!+\!\nabla_{\mu}\ln{\phi^{2}}\!=\!0\label{dep2}
\end{eqnarray}
with $R_{\mu a}^{\phantom{\mu a}a}\!=\!R_{\mu}$ and $\frac{1}{2}\varepsilon_{\mu\alpha\nu\iota}R^{\alpha\nu\iota}\!=\!B_{\mu}$ and which could be proven to be equivalent to the Dirac field equations, discussed thoroughly in \cite{Fabbri:2016laz}. The Dirac spinor field equations (\ref{D}) consist of $8$ real equations, which are as many as the $2$ vectorial equations given by the (\ref{dep1}, \ref{dep2}) above specifying all space-time derivatives of the two degrees of freedom given by YT angle and module. It is also important to notice that the YT angle, in its being the phase difference between chiral projections, must be expected to be present in the mass term, and it is, as easy to see.

As known, Maxwell and Riemann tensors encode electrodynamic and gravitational information. As such, they act as filters keeping out information of gauge and frames.

But the other hand, the gauge-invariant vector momentum and tensorial connection (\ref{P}, \ref{R}) contain the information about electrodynamics and gravity but also additional information related to the gauge and the frames, and this last type of information is not necessarily trivial despite being described by covariant objects. To see what is the information related to the gauge and the frames one should consider the conditions given by
\begin{eqnarray}
&\nabla_{\mu}P_{\nu}\!-\!\nabla_{\nu}P_{\mu}\!=\!0\label{a}\\
&\nabla_{\mu}R^{i}_{\phantom{i}j\nu}\!-\!\!\nabla_{\nu}R^{i}_{\phantom{i}j\mu}
\!\!+\!R^{i}_{\phantom{i}k\mu}R^{k}_{\phantom{k}j\nu}
\!-\!R^{i}_{\phantom{i}k\nu}R^{k}_{\phantom{k}j\mu}\!=\!0\label{b}
\end{eqnarray}
and find solutions for $P_{\mu}$ and $R^{i}_{\phantom{i}j\nu}$ that are non-zero, as this would mean that they are non-trivial and since they are tensors then they will remain such for all gauges and in all frames. However, they would contain no electrodynamic or gravitational information. The first instance is easy because (\ref{a}) is solved for non-zero gauge-invariant vector momenta of the type $P_{\mu}\!=\!\nabla_{\mu}P$ in general. As for the case of gravity things are more complicated because there does not appear to be a general solution. A special solution can however be found in spherical coordinates
\begin{eqnarray}
&g_{tt}\!=\!1\\
&g_{rr}\!=\!-1\\
&g_{\theta\theta}\!=\!-r^{2}\\
&g_{\varphi\varphi}\!=\!-r^{2}|\!\sin{\theta}|^{2}
\end{eqnarray}
with connection
\begin{eqnarray}
&\Lambda^{\theta}_{\theta r}\!=\!\frac{1}{r}\\
&\Lambda^{r}_{\theta\theta}\!=\!-r\\
&\Lambda^{\varphi}_{\varphi r}\!=\!\frac{1}{r}\\
&\Lambda^{r}_{\varphi\varphi}\!=\!-r|\!\sin{\theta}|^{2}\\
&\Lambda^{\varphi}_{\varphi\theta}\!=\!\cot{\theta}\\
&\Lambda^{\theta}_{\varphi\varphi}\!=\!-\cos{\theta}\sin{\theta}
\end{eqnarray}
by specifying to the case
\begin{eqnarray}
&u_{t}\!=\!\cosh{\alpha}\label{u1}\\
&u_{\varphi}\!=\!r\sin{\theta}\sinh{\alpha}\label{u2}
\end{eqnarray}
\begin{eqnarray}
&s_{r}\!=\!\cos{\gamma}\label{s1}\\
&s_{\theta}\!=\!r\sin{\gamma}\label{s2}
\end{eqnarray}
with $\alpha\!=\!\alpha(r,\theta)$ and $\gamma\!=\!\gamma(r,\theta)$ generic functions. Hence relations (\ref{ds},\ref{du}) can be solved for $R_{ijk}$ giving
\begin{eqnarray}
&R_{t\varphi\varphi}\!=\!R_{\varphi tt}\!=\!R_{r\theta\varphi}\!=\!R_{r\theta t}\!=\!0
\end{eqnarray}
as well as
\begin{eqnarray}
&r\sin{\theta}\partial_{\theta}\alpha\!=\!R_{t\varphi\theta}\\
&r\sin{\theta}\partial_{r}\alpha\!=\!R_{t\varphi r}\\
&-r(1\!+\!\partial_{\theta}\gamma)\!=\!R_{r\theta\theta}\\
&r\partial_{r}\gamma\!=\!R_{\theta rr}
\end{eqnarray}
linking the derivatives of the two above functions to four of the components of the $R_{ijk}$ tensor and
\begin{eqnarray}
&\!rR_{rt\varphi}\!=\!R_{t\theta\varphi}\tan{\gamma}\\
&\!\!r\sin{\theta}R_{t\theta\varphi}\!=\!(R_{\varphi\theta\varphi}
\!-\!r^{2}\cos{\theta}\sin{\theta})\tanh{\alpha}\\
&\!\!\!\!(R_{\varphi\theta\varphi}\!-\!r^{2}\sin{\theta}\cos{\theta})\tan{\gamma}\!=\!
r(R_{r\varphi\varphi}\!+\!r|\!\sin{\theta}|^{2})\\
&\!\!(R_{r\varphi\varphi}\!+\!r|\!\sin{\theta}|^{2})\tanh{\alpha}
\!=\!r\sin{\theta}R_{rt\varphi}
\end{eqnarray}
as well as
\begin{eqnarray}
&rR_{rtt}\!=\!R_{t\theta t}\tan{\gamma}\\
&r\sin{\theta}R_{t\theta t}\!=\!R_{\varphi\theta t}\tanh{\alpha}\\
&R_{\varphi\theta t}\tan{\gamma}\!=\!rR_{r\varphi t}\\
&R_{r\varphi t}\tanh{\alpha}\!=\!r\sin{\theta}R_{rtt}
\end{eqnarray}
and
\begin{eqnarray}
&rR_{rtr}\!=\!R_{t\theta r}\tan{\gamma}\\
&r\sin{\theta}R_{t\theta r}\!=\!R_{\varphi\theta r}\tanh{\alpha}\\
&R_{\varphi\theta r}\tan{\gamma}\!=\!rR_{r\varphi r}\\
&R_{r\varphi r}\tanh{\alpha}\!=\!r\sin{\theta}R_{rtr}
\end{eqnarray}
with
\begin{eqnarray}
&rR_{rt\theta}\!=\!R_{t\theta\theta}\tan{\gamma}\\
&r\sin{\theta}R_{t\theta\theta}\!=\!R_{\varphi\theta\theta}\tanh{\alpha}\\
&R_{\varphi\theta\theta}\tan{\gamma}\!=\!rR_{r\varphi\theta}\\
&R_{r\varphi\theta}\tanh{\alpha}\!=\!r\sin{\theta}R_{rt\theta}
\end{eqnarray}
grouped in four independent blocks each with four interlinked relations. Thus a working hypothesis might be to look for a solution in which we can set to zero some block while leaving different from zero others, such as 
\begin{eqnarray}
&R_{trr}\!=\!R_{t\theta r}\!=\!R_{\varphi rr}\!=\!R_{\varphi\theta r}\!=\!0\\
&R_{rt\theta}\!=\!R_{t\theta\theta}\!=\!R_{r\varphi\theta}\!=\!R_{\varphi\theta\theta}\!=\!0\\
&R_{t\theta\varphi}\!=\!R_{tr\varphi}\!=\!0
\end{eqnarray}
with
\begin{eqnarray}
&R_{r\varphi\varphi}\!=\!-r|\!\sin{\theta}|^{2}\\
&R_{\theta\varphi\varphi}\!=\!-r^{2}\cos{\theta}\sin{\theta}
\end{eqnarray}
and
\begin{eqnarray}
&R_{rtt}\!=\!-2\varepsilon\sinh{\alpha}\sin{\gamma}\\
&R_{\varphi rt}\!=\!2\varepsilon r\sin{\theta}\cosh{\alpha}\sin{\gamma}\\
&R_{\theta tt}\!=\!2\varepsilon r\sinh{\alpha}\cos{\gamma}\\
&R_{\varphi\theta t}\!=\!-2\varepsilon r^{2}\sin{\theta}\cosh{\alpha}\cos{\gamma}
\end{eqnarray}
with $\varepsilon$ being a generic constant, which can be interpreted as an integration constant since it comes from having set the Riemann curvature tensor to be zero identically \cite{Fabbri:2019kfr}.

Therefore, we have shown that it is indeed possible to have zero Riemann and Maxwell tensors, thus no gravity and electrodynamics, but still have non-vanishing covariant objects that as such contain information related only to frames and gauge, but that are non-trivial and cannot be removed with choices of frames and gauge. That this had to be the case is also clear from the fact that setting gauge-invariant vector momenta or tensorial connections to zero in general leads to unwanted consequences \cite{Fabbri:2017pwp}.

The tensorial connection is just the gravitational analogue of what the gauge-invariant vector momentum is in electrodynamics. However, the tensorial connection has a much richer structure since it can be decomposed into the vector trace $R_{a}$ and the axial-vector dual $B_{a}$ with a non-completely antisymmetric irreducible part accounting for the rest of the information on the degrees of freedom \cite{Fabbri:2020ypd}.

We notice that $B_{a}$ is an axial-vector that contains the same information of the connection and it also possesses the same mass dimension of the connection itself.
\section{Re-normalizable Chern-Simons Stelle Gravity}
In the introduction we have discussed that a first way to have a homogeneous Chern-Simons correction to Einstein gravity is to find the square-curvature leading term to be added to the already-mentioned square-curvature Chern-Simons type of gravitational topological current.

The most general square-curvature leading term is
\begin{eqnarray}
&\mathscr{L}\!=\!XR^{\alpha\pi\sigma\nu}R_{\alpha\pi\sigma\nu}
\!+\!YR^{\sigma\nu}R_{\sigma\nu}\!+\!ZR^{2}
\end{eqnarray}
where $R_{\alpha\pi\sigma\nu}$ is the Riemann tensor, $R_{\pi\nu}\!=\!R^{\alpha}_{\phantom{\alpha}\pi\alpha\nu}$ is the Ricci tensor and $R\!=\!R^{\alpha}_{\phantom{\alpha}\alpha}$ is the Ricci scalar. Nonetheless, the Gauss-Bonnet identity tells that the square-Riemann term can always be written as a combination of the two square-Ricci terms up to a divergence. So there is no loss of generality in setting $X$ to zero and considering
\begin{eqnarray}
&\mathscr{L}\!=\!YR^{\sigma\nu}R_{\sigma\nu}\!+\!ZR^{2}
\end{eqnarray}
as the square-curvature Lagrangian \cite{Stelle:1977ry}. This Lagrangian is re-normalizable \cite{Stelle:1976gc}. We call it Stelle Lagrangian.

To this we have to add the term
\begin{eqnarray}
&\mathscr{L}\!=\!KbR^{pq\sigma\nu}R^{ac}_{\phantom{ac}\sigma\nu}\varepsilon_{pqac}
\end{eqnarray}
with $b$ a general pseudo-scalar \cite{Jackiw:2003pm}. The square-curvature term can be written as $R^{pq\sigma\nu}R^{ac}_{\phantom{ac}\sigma\nu}\varepsilon_{pqac}\!=\!\nabla_{\mu}K^{\mu}$ as well known, but now with the tools developed in the previous section we can also see that the axial-vector is explicitly given by $K^{\mu}\!=\!-4\varepsilon^{\mu\nu\sigma\rho}
(R^{a}_{\phantom{a}c\nu}\!\nabla_{\sigma}R^{c}_{\phantom{c}a\rho}
\!+\!\frac{4}{3}R^{c}_{\phantom{c}a\nu}R^{a}_{\phantom{a}v\sigma}R^{v}_{\phantom{v}c\rho})$ in terms of the tensorial connection alone. And because the Yvon-Takabayashi angle is a pseudo-scalar of zero mass dimension, we will consider $b$ to be $\beta$ so that
\begin{eqnarray}
&\mathscr{L}\!=\!K\beta R^{pq\sigma\nu}R^{ac}_{\phantom{ac}\sigma\nu}\varepsilon_{pqac}
\end{eqnarray}
is a square-curvature of $4$ mass dimension, and the topological term has the same properties of the leading term.

The most general Lagrangian is given by
\begin{eqnarray}
&\mathscr{L}\!=\!YR^{\sigma\nu}R_{\sigma\nu}\!+\!ZR^{2}
\!+\!K\beta R^{pq\sigma\nu}R^{ac}_{\phantom{ac}\sigma\nu}\varepsilon_{pqac}
\end{eqnarray}
which is in fact homogeneous and re-normalizable, and it is the Lagrangian of the gravitational sector. Then
\begin{eqnarray}
\nonumber
&\mathscr{L}\!=\!YR^{\sigma\nu}R_{\sigma\nu}\!+\!ZR^{2}
\!+\!K\beta R^{pq\sigma\nu}R^{ac}_{\phantom{ac}\sigma\nu}\varepsilon_{pqac}-\\
&-i\overline{\psi}\boldsymbol{\gamma}^{\mu}\boldsymbol{\nabla}_{\mu}\psi\!+\!m\overline{\psi}\psi
\end{eqnarray}
or in polar form
\begin{eqnarray}
\nonumber
&\mathscr{L}\!=\!YR^{\sigma\nu}R_{\sigma\nu}\!+\!ZR^{2}
\!+\!K\beta R^{pq\sigma\nu}R^{ac}_{\phantom{ac}\sigma\nu}\varepsilon_{pqac}-\\
&-\phi^{2}[s^{\mu}(\nabla_{\mu}\beta\!+\!B_{\mu})\!+\!2u^{\mu}P_{\mu}\!-\!2m\cos{\beta}]
\end{eqnarray}
is the Lagrangian of gravitational and material sectors.

The variation of this Lagrangian is performed by employing the usual method, although here the condition of torsionlessness means that the variation of the connection is given in terms of the variation of the metric itself with the result that the gravitational field equations are
\begin{eqnarray}
\nonumber
&Y\nabla^{2}R_{\mu\nu}\!+\!\frac{1}{2}(4Z\!+\!Y)\nabla^{2}Rg_{\mu\nu}
\!-\!(2Z\!+\!Y)\nabla_{\mu}\nabla_{\nu}R+\\
\nonumber
&+2YR_{\mu\rho\nu\sigma}R^{\rho\sigma}\!-\!\frac{1}{2}YR_{\alpha\rho}R^{\alpha\rho}g_{\mu\nu}+\\
\nonumber
&+2ZRR_{\mu\nu}\!-\!\frac{1}{2}ZR^{2}g_{\mu\nu}+\\
\nonumber
&+2K\nabla^{\rho}(\nabla^{\alpha}\beta R_{\rho\mu}^{\phantom{\rho\mu}\pi\sigma}\varepsilon_{\alpha\nu\pi\sigma}
\!+\!\nabla^{\alpha}\beta R_{\rho\nu}^{\phantom{\rho\nu}\pi\sigma}\varepsilon_{\alpha\mu\pi\sigma})=\\
&=\frac{i}{8}(\overline{\psi}\boldsymbol{\gamma}_{\nu}\boldsymbol{\nabla}_{\mu}\psi
\!-\!\boldsymbol{\nabla}_{\mu}\overline{\psi}\boldsymbol{\gamma}_{\nu}\psi
\!+\!\overline{\psi}\boldsymbol{\gamma}_{\mu}\boldsymbol{\nabla}_{\nu}\psi
\!-\!\boldsymbol{\nabla}_{\nu}\overline{\psi}\boldsymbol{\gamma}_{\mu}\psi)
\label{ee}
\end{eqnarray}
while the variation with respect to the spinor field furnishes the material field equations
\begin{eqnarray}
&\!\!\!\!\!\!\!\!i\boldsymbol{\gamma}^{\mu}\boldsymbol{\nabla}_{\mu}\psi
\!-\!\frac{K}{2}R^{pq\sigma\nu}R^{ac}_{\phantom{ac}\sigma\nu}\varepsilon_{pqac}\phi^{-2}e^{i\boldsymbol{\pi}\beta}i\boldsymbol{\pi}\psi\!-\!m\psi\!=\!0
\end{eqnarray}
or respectively in polar form
\begin{eqnarray}
\nonumber
&Y\nabla^{2}R_{\mu\nu}\!+\!\frac{1}{2}(4Z\!+\!Y)\nabla^{2}Rg_{\mu\nu}
\!-\!(2Z\!+\!Y)\nabla_{\mu}\nabla_{\nu}R+\\
\nonumber
&+2YR_{\mu\rho\nu\sigma}R^{\rho\sigma}\!-\!\frac{1}{2}YR_{\alpha\rho}R^{\alpha\rho}g_{\mu\nu}+\\
\nonumber
&+2ZRR_{\mu\nu}\!-\!\frac{1}{2}ZR^{2}g_{\mu\nu}+\\
\nonumber
&+2K\nabla^{\rho}(\nabla^{\alpha}\beta R_{\rho\mu}^{\phantom{\rho\mu}\pi\sigma}\varepsilon_{\alpha\nu\pi\sigma}
\!+\!\nabla^{\alpha}\beta R_{\rho\nu}^{\phantom{\rho\nu}\pi\sigma}\varepsilon_{\alpha\mu\pi\sigma})=\\
\nonumber
&=\frac{1}{2}\phi^{2}(u_{\nu}P_{\mu}\!+\!u_{\mu}P_{\nu}
\!+\!\frac{1}{2}s_{\nu}\nabla_{\mu}\beta\!+\!\frac{1}{2}s_{\mu}\nabla_{\nu}\beta-\\
&-\frac{1}{4}R^{ij}_{\phantom{ij}\mu}\varepsilon_{\nu ijk}s^{k}
\!-\!\frac{1}{4}R^{ij}_{\phantom{ij}\nu}\varepsilon_{\mu ijk}s^{k})
\end{eqnarray}
and
\begin{eqnarray}
&-2P^{\iota}u_{[\iota}s_{\mu]}\!+\!B_{\mu}\!+\!\nabla_{\mu}\beta\!+\!2s_{\mu}m\cos{\beta}\!=\!0\\
\nonumber
&-2P^{\rho}u^{\nu}s^{\alpha}\varepsilon_{\mu\rho\nu\alpha}\!+\!R_{\mu}
\!-\!s_{\mu}KR^{pq\sigma\nu}R^{ac}_{\phantom{ac}\sigma\nu}\varepsilon_{pqac}\phi^{-2}+\\
&+\nabla_{\mu}\ln{\phi^{2}}\!+\!2s_{\mu}m\sin{\beta}\!=\!0
\end{eqnarray}
as the full set of gravity and matter field equations.

Taking the divergence of field equations (\ref{ee}) and using the Jacobi-Bianchi cyclic identities gives the constraints
\begin{eqnarray}
\nonumber
&\!\!-K\nabla_{\nu}\beta R^{pq\alpha\sigma}R^{ij}_{\phantom{ij}\alpha\sigma}\varepsilon_{ijpq}=\\
&\!\!\!\!=\frac{i}{4}\nabla^{\mu}
(\overline{\psi}\boldsymbol{\gamma}_{\nu}\!\boldsymbol{\nabla}_{\mu}\psi
\!-\!\!\boldsymbol{\nabla}_{\mu}\overline{\psi}\boldsymbol{\gamma}_{\nu}\psi
\!+\!\overline{\psi}\boldsymbol{\gamma}_{\mu}\!\boldsymbol{\nabla}_{\nu}\psi
\!-\!\!\boldsymbol{\nabla}_{\nu}\overline{\psi}\boldsymbol{\gamma}_{\mu}\psi)
\label{constraint}
\end{eqnarray}
showing that the divergence of the symmetric energy density tensor is not zero. Nevertheless, computing it as
\begin{eqnarray}
\nonumber
&\frac{i}{4}\nabla^{\mu}
(\overline{\psi}\boldsymbol{\gamma}_{\nu}\boldsymbol{\nabla}_{\mu}\psi
\!-\!\boldsymbol{\nabla}_{\mu}\overline{\psi}\boldsymbol{\gamma}_{\nu}\psi
\!+\!\overline{\psi}\boldsymbol{\gamma}_{\mu}\boldsymbol{\nabla}_{\nu}\psi
\!-\!\boldsymbol{\nabla}_{\nu}\overline{\psi}\boldsymbol{\gamma}_{\mu}\psi)=\\
\nonumber
&=\frac{i}{4}[\ \overline{\psi}\boldsymbol{\gamma}_{\nu}\boldsymbol{\nabla}^{2}\psi
\!-\!\boldsymbol{\nabla}^{2}\overline{\psi}\boldsymbol{\gamma}_{\nu}\psi+\\
\nonumber
&+(\boldsymbol{\nabla}^{\mu}\overline{\psi}\boldsymbol{\gamma}_{\mu})
\boldsymbol{\nabla}_{\nu}\psi
\!+\!\overline{\psi}\boldsymbol{\gamma}^{\mu}
[\boldsymbol{\nabla}_{\mu},\boldsymbol{\nabla}_{\nu}]\psi+\\
\nonumber
&+\overline{\psi}\boldsymbol{\nabla}_{\nu}
(\boldsymbol{\gamma}^{\mu}\boldsymbol{\nabla}_{\mu}\psi)
\!-\!\boldsymbol{\nabla}_{\nu}
(\boldsymbol{\nabla}_{\mu}\overline{\psi}\boldsymbol{\gamma}^{\mu})\psi-\\
&-[\boldsymbol{\nabla}_{\mu},\boldsymbol{\nabla}_{\nu}]\overline{\psi}
\boldsymbol{\gamma}^{\mu}\psi
\!-\!\boldsymbol{\nabla}_{\nu}\overline{\psi}(\boldsymbol{\gamma}_{\mu}
\boldsymbol{\nabla}^{\mu}\psi)]
\end{eqnarray}
and repeatedly employing the Dirac equations gives
\begin{eqnarray}
\nonumber
&\frac{i}{4}[\ \overline{\psi}\boldsymbol{\gamma}_{\nu}\boldsymbol{\nabla}^{2}\psi
\!-\!\boldsymbol{\nabla}^{2}\overline{\psi}\boldsymbol{\gamma}_{\nu}\psi+\\
\nonumber
&+(\boldsymbol{\nabla}^{\mu}\overline{\psi}\boldsymbol{\gamma}_{\mu})
\boldsymbol{\nabla}_{\nu}\psi
\!+\!\overline{\psi}\boldsymbol{\gamma}^{\mu}
[\boldsymbol{\nabla}_{\mu},\boldsymbol{\nabla}_{\nu}]\psi+\\
\nonumber
&+\overline{\psi}\boldsymbol{\nabla}_{\nu}
(\boldsymbol{\gamma}^{\mu}\boldsymbol{\nabla}_{\mu}\psi)
\!-\!\boldsymbol{\nabla}_{\nu}
(\boldsymbol{\nabla}_{\mu}\overline{\psi}\boldsymbol{\gamma}^{\mu})\psi-\\
\nonumber
&-[\boldsymbol{\nabla}_{\mu},\boldsymbol{\nabla}_{\nu}]\overline{\psi}
\boldsymbol{\gamma}^{\mu}\psi
\!-\!\boldsymbol{\nabla}_{\nu}\overline{\psi}(\boldsymbol{\gamma}_{\mu}
\boldsymbol{\nabla}^{\mu}\psi)]=\\
&=\frac{K}{2}\overline{\psi}\boldsymbol{\nabla}_{\nu}
(R^{pq\sigma\alpha}R^{ac}_{\phantom{ac}\sigma\alpha}\varepsilon_{pqac}\phi^{-2}
e^{i\boldsymbol{\pi}\beta}i\boldsymbol{\pi})\psi
\end{eqnarray}
which can eventually be easily turned into
\begin{eqnarray}
\nonumber
&\frac{K}{2}\overline{\psi}\boldsymbol{\nabla}_{\nu}
(R^{pq\sigma\alpha}R^{ac}_{\phantom{ac}\sigma\alpha}\varepsilon_{pqac}\phi^{-2}
e^{i\boldsymbol{\pi}\beta}i\boldsymbol{\pi})\psi=\\
&=-KR^{pq\sigma\alpha}R^{ac}_{\phantom{ac}\sigma\alpha}\varepsilon_{pqac}\nabla_{\nu}\beta
\end{eqnarray}
as clear. Putting together the last three expressions gives
\begin{eqnarray}
\nonumber
&\frac{i}{4}\nabla^{\mu}
(\overline{\psi}\boldsymbol{\gamma}_{\nu}\boldsymbol{\nabla}_{\mu}\psi
\!-\!\boldsymbol{\nabla}_{\mu}\overline{\psi}\boldsymbol{\gamma}_{\nu}\psi
\!+\!\overline{\psi}\boldsymbol{\gamma}_{\mu}\boldsymbol{\nabla}_{\nu}\psi
\!-\!\boldsymbol{\nabla}_{\nu}\overline{\psi}\boldsymbol{\gamma}_{\mu}\psi)=\\
&=-KR^{pq\sigma\alpha}R^{ac}_{\phantom{ac}\sigma\alpha}\varepsilon_{pqac}\nabla_{\nu}\beta
\end{eqnarray}
to be compared against (\ref{constraint}) above. In doing so, we finally see that there is no constraint developed in the end.

This is important since in \cite{Jackiw:2003pm} the authors state that the theory must be restricted to have $R^{pq\sigma\nu}R^{ac}_{\phantom{ac}\sigma\nu}\varepsilon_{pqac}\!\equiv\!0$ in order for the divergencelessness of the symmetric energy density tensor to hold. Such a restriction is not necessary as the divergencelessness of the symmetric energy density tensor does not need to be implemented in the first place due to the fact that the $\beta R^{pq\sigma\nu}R^{ac}_{\phantom{ac}\sigma\nu}\varepsilon_{pqac}$ is a potential energy of interaction between matter and the space-time.

Lack of conservation of energy simply means that there is an energy flux from/to matter, which does not have to be zero so long as it is exactly compensated by the energy flux to/from the space-time, and here we proved it is.

We will see however that this is not always true.
\section{Least-Derivative Chern-Simons Hilbert Gravity}
In the introduction, we have discussed about the necessity to have an object $K_{a}$ being an axial-vector and built in terms of the connection. And in the previous section, we have seen that the axial-vector dual of the tensorial connection has such features. As such $B_{a}$ seems the perfect candidate for the topological current we are seeking.

To the least-order derivative the leading term is
\begin{eqnarray}
&\mathscr{L}\!=\!R
\end{eqnarray}
which is the least-order derivative Lagrangian. Obviously this is the very well known usual Hilbert Lagrangian.

The least-order derivative topological term can now be added straightforwardly as it has to be of the form
\begin{eqnarray}
&\mathscr{L}\!=\!kb\nabla_{\mu}B^{\mu}
\end{eqnarray}
with $b$ generic pseudo-scalar. However, if we want to keep homogeneity, and since the gravitational Lagrangian has $2$ mass dimension, then also this term must have $2$ mass dimension, and the only way we have to do this is to take $b$ to be a pseudo-scalar of zero mass dimension, which can only be the Yvon-Takabayashi angle, so that
\begin{eqnarray}
&\mathscr{L}\!=\!k\beta\nabla_{\mu}B^{\mu}
\end{eqnarray}
is the only option for the least-order topological term.

Altogether we have 
\begin{eqnarray}
&\mathscr{L}\!=\!R\!+\!k\beta\nabla_{\mu}B^{\mu}
\end{eqnarray}
which is homogeneous and least-order derivative, as the Lagrangian for gravity. Therefore we have that
\begin{eqnarray}
&\mathscr{L}\!=\!R\!+\!k\beta\nabla_{\mu}B^{\mu}
\!-\!i\overline{\psi}\boldsymbol{\gamma}^{\mu}\boldsymbol{\nabla}_{\mu}\psi
\!+\!m\overline{\psi}\psi
\end{eqnarray}
or in polar form
\begin{eqnarray}
\nonumber
&\mathscr{L}\!=\!R\!+\!k\beta\nabla_{\mu}B^{\mu}\!-\!\phi^{2}[s^{\mu}(\nabla_{\mu}\beta\!+\!B_{\mu})+\\
&+2u^{\mu}P_{\mu}\!-\!2m\cos{\beta}]
\end{eqnarray}
for the gravitational and the material sectors together.

The variation of this Lagrangian then gives
\begin{eqnarray}
\nonumber
&R^{\nu\sigma}\!-\!\frac{1}{2}g^{\nu\sigma}R
\!+\!\frac{k}{2}[-\beta\nabla_{\mu} B^{\mu}g^{\nu\sigma}+\\
\nonumber
&+\frac{1}{2}\nabla_{\mu}\beta
(\varepsilon^{\mu\sigma\alpha\eta}R^{\nu}_{\phantom{\nu}\alpha\eta}
\!+\!\varepsilon^{\mu\nu\alpha\eta}R^{\sigma}_{\phantom{\sigma}\alpha\eta}
\!-\!2B^{\mu}g^{\nu\sigma})]=\\
&\!\!\!\!=\frac{i}{8}(\overline{\psi}\boldsymbol{\gamma}^{\nu}\!\boldsymbol{\nabla}^{\sigma}\psi
\!-\!\!\boldsymbol{\nabla}^{\sigma}\overline{\psi}\boldsymbol{\gamma}^{\nu}\psi
\!+\!\overline{\psi}\boldsymbol{\gamma}^{\sigma}\!\boldsymbol{\nabla}^{\nu}\psi
\!-\!\!\boldsymbol{\nabla}^{\nu}\overline{\psi}\boldsymbol{\gamma}^{\sigma}\psi)
\label{e}
\end{eqnarray}
and
\begin{eqnarray}
&\!\!\!\!i\boldsymbol{\gamma}^{\mu}\boldsymbol{\nabla}_{\mu}\psi
\!-\!\frac{k}{2}\nabla\!\cdot\!B\phi^{-2}e^{i\boldsymbol{\pi}\beta}i\boldsymbol{\pi}\psi
\!-\!m\psi\!=\!0
\end{eqnarray}
or respectively in polar form
\begin{eqnarray}
\nonumber
&R^{\nu\sigma}\!-\!\frac{1}{2}g^{\nu\sigma}R
\!+\!\frac{k}{2}[-\beta\nabla_{\mu} B^{\mu}g^{\nu\sigma}+\\
\nonumber
&+\frac{1}{2}\nabla_{\mu}\beta
(\varepsilon^{\mu\sigma\alpha\eta}R^{\nu}_{\phantom{\nu}\alpha\eta}
\!+\!\varepsilon^{\mu\nu\alpha\eta}R^{\sigma}_{\phantom{\sigma}\alpha\eta}
\!-\!2B^{\mu}g^{\nu\sigma})]=\\
\nonumber
&=\frac{1}{2}\phi^{2}(u^{\nu}P^{\sigma}\!+\!u^{\sigma}P^{\nu}
\!+\!\frac{1}{2}s^{\nu}\nabla^{\sigma}\beta\!+\!\frac{1}{2}s^{\sigma}\nabla^{\nu}\beta-\\
&-\frac{1}{4}R_{ij}^{\phantom{ij}\sigma}\varepsilon^{\nu ijk}s_{k}
\!-\!\frac{1}{4}R_{ij}^{\phantom{ij}\nu}\varepsilon^{\sigma ijk}s_{k})
\label{e1}
\end{eqnarray}
and
\begin{eqnarray}
&-2P^{\iota}u_{[\iota}s_{\mu]}\!+\!B_{\mu}\!+\!\nabla_{\mu}\beta
\!+\!2s_{\mu}m\cos{\beta}\!=\!0\label{d1}\\
\nonumber
&-2P^{\rho}u^{\nu}s^{\alpha}\varepsilon_{\mu\rho\nu\alpha}\!+\!R_{\mu}
\!-\!s_{\mu}k\nabla\!\cdot\!B\phi^{-2}+\\
&+\nabla_{\mu}\ln{\phi^{2}}\!+\!2s_{\mu}m\sin{\beta}\!=\!0\label{d2}
\end{eqnarray}
as the full set of gravity and matter field equations.

Taking the divergence of field equations (\ref{e}) and using the Jacobi-Bianchi cyclic identities yields that
\begin{eqnarray}
\nonumber
&\!\!k\nabla_{\nu}[-\beta\nabla_{\mu} B^{\mu}g^{\nu\sigma}+\\
\nonumber
&\!\!\!\!+\frac{1}{2}\nabla_{\mu}\beta
(\varepsilon^{\mu\sigma\alpha\eta}R^{\nu}_{\phantom{\nu}\alpha\eta}
\!+\!\varepsilon^{\mu\nu\alpha\eta}R^{\sigma}_{\phantom{\sigma}\alpha\eta}
\!-\!2B^{\mu}g^{\nu\sigma})]=\\
&\!\!\!\!\!\!\!\!\!\!\!\!=\frac{i}{4}\nabla_{\nu}(\overline{\psi}\boldsymbol{\gamma}^{\nu}\!\boldsymbol{\nabla}^{\sigma}\psi
\!-\!\!\boldsymbol{\nabla}^{\sigma}\overline{\psi}\boldsymbol{\gamma}^{\nu}\psi
\!+\!\overline{\psi}\boldsymbol{\gamma}^{\sigma}\!\boldsymbol{\nabla}^{\nu}\psi
\!-\!\!\boldsymbol{\nabla}^{\nu}\overline{\psi}\boldsymbol{\gamma}^{\sigma}\psi)
\label{restriction}
\end{eqnarray}
showing that the divergence of the symmetric energy density tensor is not zero, similarly as before. However, now by following the same strategy we would obtain that
\begin{eqnarray}
\nonumber
&\frac{i}{4}\nabla_{\nu}(\overline{\psi}\boldsymbol{\gamma}^{\nu}\!\boldsymbol{\nabla}^{\sigma}\psi
\!-\!\!\boldsymbol{\nabla}^{\sigma}\overline{\psi}\boldsymbol{\gamma}^{\nu}\psi
\!+\!\overline{\psi}\boldsymbol{\gamma}^{\sigma}\!\boldsymbol{\nabla}^{\nu}\psi
\!-\!\!\boldsymbol{\nabla}^{\nu}\overline{\psi}\boldsymbol{\gamma}^{\sigma}\psi)=\\
&=-k\nabla\!\cdot\!B\nabla^{\sigma}\beta
\end{eqnarray}
to be compared against (\ref{restriction}) above. In doing so, we get 
\begin{eqnarray}
\nonumber
&\nabla_{\mu}\nabla_{\nu}\beta\varepsilon^{\mu\sigma\alpha\eta}R^{\nu}_{\phantom{\nu}\alpha\eta}
\!-\!2\nabla^{\mu}\nabla^{\sigma}\beta B_{\mu}+\\
\nonumber
&+\nabla_{\mu}\beta\varepsilon^{\mu\sigma\alpha\eta}
\nabla_{\nu}R^{\nu}_{\phantom{\nu}\alpha\eta}
\!+\!\nabla_{\mu}\beta\varepsilon^{\mu\nu\alpha\eta}
\nabla_{\nu}R^{\sigma}_{\phantom{\sigma}\alpha\eta}-\\
&-2\nabla_{\mu}\beta \nabla^{\sigma}B^{\mu}
\!-\!2\beta\nabla^{\sigma}\nabla^{\mu}B_{\mu}\!=\!0
\label{r}
\end{eqnarray}
with the restriction still holding. Because of the different dependence on the differential structure of the YT angle, there is no way to have this solved in general. Therefore, in the present case we remain with a true restriction that has to be imposed on the structure of the background.

The reason for this occurrence is that for determining the conservation laws we have to use field equations which for the tensorial connection $R_{ij\alpha}$ do not exist \cite{Fabbri:2020ypd}. In fact, not only the $R_{ij\alpha}$ have no dynamical equations coupling them to sources, but they also have non-local properties and in particular, they do not need to vanish at infinity.

The field equations (\ref{e}) can now be studied to see the effects on the background. Effects on the curvature have to be expected, but now effects on the topological sector should be expected as well. To screen the curvature, and isolate the effects on topology, we should consider the flat space-time, but endowed with some non-trivial tensorial connection. Luckily, we already have it, as it is given by the components above. In absence of electrodynamics $P_{\mu}$ consists of a pure gauge, and the information encoded in $P_{\mu}$ can always be written as $P_{\mu}\!=\!(m,\vec{0})$ all throughout.

In this case field equations (\ref{e1}) reduce to
\begin{eqnarray}
\nonumber
&k[\frac{1}{2}\nabla_{\mu}\beta
(\varepsilon^{\mu\sigma\alpha\eta}R^{\nu}_{\phantom{\nu}\alpha\eta}
\!+\!\varepsilon^{\mu\nu\alpha\eta}R^{\sigma}_{\phantom{\sigma}\alpha\eta})-\\
\nonumber
&-\nabla_{\mu}(\beta B^{\mu})g^{\nu\sigma}]\!=\!\phi^{2}(u^{\nu}P^{\sigma}\!+\!u^{\sigma}P^{\nu}+\\
\nonumber
&+\frac{1}{2}s^{\nu}\nabla^{\sigma}\beta\!+\!\frac{1}{2}s^{\sigma}\nabla^{\nu}\beta-\\
&-\frac{1}{4}R_{ij}^{\phantom{ij}\sigma}\varepsilon^{\nu ijk}s_{k}
\!-\!\frac{1}{4}R_{ij}^{\phantom{ij}\nu}\varepsilon^{\sigma ijk}s_{k})
\end{eqnarray}
while the Dirac field equations (\ref{d1}, \ref{d2}) reduce to
\begin{eqnarray}
&-2P^{\iota}u_{[\iota}s_{\mu]}\!+\!B_{\mu}\!+\!\nabla_{\mu}\beta\!+\!2s_{\mu}m\cos{\beta}\!=\!0\\
\nonumber
&-2P^{\rho}u^{\nu}s^{\alpha}\varepsilon_{\mu\rho\nu\alpha}\!+\!R_{\mu}
\!-\!s_{\mu}k\nabla\!\cdot\!B\phi^{-2}+\\
&+\nabla_{\mu}\ln{\phi^{2}}\!+\!2s_{\mu}m\sin{\beta}\!=\!0
\end{eqnarray}
with the expression of the tensorial connection that has to be plugged in. These field equations have to be studied in specific situations, although finding an exact solution is quite generally not an easy task to be accomplished.

To simplify the problem, we consider that, in order to study the topological features at infinity, we are allowed to take into account the macroscopic approximation.

In this case, all internal structures are negligible and with them also the spin contributions. We can thus take the momentum to be $P_{\alpha}\!=\!m\cos{\beta}u_{\alpha}$ \cite{Fabbri:2020ypd}, so to write
\begin{eqnarray}
\nonumber
&k[\frac{1}{2}\nabla_{\mu}\beta
(\varepsilon^{\mu\sigma\alpha\eta}R^{\nu}_{\phantom{\nu}\alpha\eta}
\!+\!\varepsilon^{\mu\nu\alpha\eta}R^{\sigma}_{\phantom{\sigma}\alpha\eta})-\\
&-\nabla_{\mu}(\beta B^{\mu})g^{\nu\sigma}]\!\approx\!2\phi^{2}m\cos{\beta}u^{\nu}u^{\sigma}
\end{eqnarray}
so that in the co-moving frame only the time-time component would account for a coupling to the material distribution. This will be eventually given by
\begin{eqnarray}
&-\nabla_{\mu}(k\beta B^{\mu})\!\approx\!2\phi^{2}m\cos{\beta}
\end{eqnarray}
where on the right-hand side we have the mass density.

Integrating over the whole volume and considering the normalization for the matter distribution we may write 
\begin{eqnarray}
\oint_{\partial V}\beta B_{\mu}dS^{\mu}\!\approx\!-m/k\!\neq\!0
\end{eqnarray}
where $dS^{\mu}$ is the element of surface $S\!=\!\partial V$ constituting the border of the volume $V$ of integration. This condition clearly shows that the vector $\beta B_{\mu}$ can never go to zero.

Therefore, the tensorial connection, as well as the YT angle, display some topological features at infinity.
\section{Final Comparison}
We have studied the re-normalizable Chern-Simons extension of Stelle gravity, described by the Lagrangian
\begin{eqnarray}
&\mathscr{L}\!=\!YR^{\sigma\nu}R_{\sigma\nu}\!+\!ZR^{2}
\!+\!K\beta R^{pq\sigma\nu}R^{ac}_{\phantom{ac}\sigma\nu}\varepsilon_{pqac}
\label{S}
\end{eqnarray}
and the least-order derivative Chern-Simons extension of the Hilbert gravity, described by the Lagrangian
\begin{eqnarray}
&\mathscr{L}\!=\!R\!+\!k\beta\nabla_{\mu}B^{\mu}
\label{H}
\end{eqnarray}
and it is now time for a comparison of the two.

In analogy with one another, the two corrections consist of an interaction between the Yvon-Takabayashi angle and the background, and therefore we should expect an energy exchange between these two. So such an energy exchange would entail the failure of the conservation law for the energy density of matter allowing only the conservation law for the energy density of the system of matter and space-time taken together. Differently from the usual procedure of constraining the structure of the space-time in order to salvage the conservation of energy for matter, here we have demonstrated that there is conservation of the symmetric energy density for the full matter-gravity system. And in the present paper we proved that this is the case, for the re-normalizable Chern-Simons extension of Stelle gravity. Nevertheless we also proved that this is not the case for the least-order derivative Chern-Simons extension of the Hilbert gravity since we got (\ref{r}) which is not verified identically. This accounts for what we see as the main difference between these two extensions.

We believe that the reason for this difference is to be sought in the fact that in the re-normalizable instance the higher-order derivative structure of the correction allows it to be written entirely in terms of the curvature tensor while in the least-order derivative instance the correction is not written in terms of the curvature but in terms of the tensorial connection. The difference between the curvature and tensorial connection is that while the former contains only information about gravity, the latter contains information about both gravity and inertial contributions. The first case is entirely physical, the correction is fully dynamically determined and hence the energy of the total system is conserved. The second case is physical only in the gravitational information, so the correction is dynamically determined only for gravitation and thus the energy non-conservation of matter is compensated by the energy non-conservation of gravity, but there the inertial information is not physical, it does not have a field equation and it is only by having it constrained that the full conservation of the energy density is ensured eventually.

The tensorial connection may contain information that does not necessarily vanish at infinity while the gravitational field has always to vanish at the boundary of the space-time. This means that there is more information on large scale structures of space-time in (\ref{H}) than (\ref{S}).
\section{Combination}
It is important to highlight that in what we have done above, we kept the two theories separated for the clearest comparison. Nonetheless, there is no other reason to keep the two theories separated, and they can be joined into
\begin{eqnarray}
\nonumber
&\mathscr{L}\!=\!YR^{\sigma\nu}R_{\sigma\nu}\!+\!ZR^{2}\!+\!R\!+\!2\Lambda+\\
&+\beta(KR^{pq\sigma\nu}R^{ac}_{\phantom{ac}\sigma\nu}\varepsilon_{pqac}
\!+\!k\nabla_{\mu}B^{\mu})
\end{eqnarray}
where also the cosmological constant has been added, and where the Yvon-Takabayashi angle has been factored out the topological term. This Lagrangian would be the most complete, although no longer least-order derivative, and with the mass dimension needed for re-normalizability.

So a natural question would now be whether or not it would actually be a re-normalizable action after all.

We hope that better physicists might answer to such a question strengthening the validity of this theory.
\section{Conclusion}
In this paper, we have discussed in what way it is possible to exploit the polar form of spinor fields, and hence the tensorial connection, to construct the re-normalizable higher-derivative CS extension of square-curvature Stelle Lagrangian and the least-derivative CS extension of the linear-curvature Hilbert Lagrangian. Then, we discussed the analogies and the differences of the two theories.

By combining the two and allowing for the cosmological constant would give the mass-dimension four topological gravitation in the most complete form.

\end{document}